\documentclass[final]{article}
\usepackage{graphicx}
\usepackage{graphics}
\usepackage[T1]{fontenc}
\usepackage{colortbl}
\usepackage{amsmath}
\usepackage{amsfonts}
\usepackage{amssymb}
\usepackage{array}
\usepackage{natbib}
\usepackage[left=33mm,right=22mm]{geometry}
\title{Missing data in a stochastic Dollo model for cognate data, and its application to the dating of Proto-Indo-European}
\author{Robin J. Ryder and Geoff K. Nicholls \\ Department of Statistics, University of Oxford, UK}

\newenvironment{newsec}{}{}

\begin{document}
\maketitle
\begin{abstract}
\citet{nicholls2006dat} describe a phylogenetic model for trait data. They use their model to estimate branching times on
Indo-European language trees from lexical data. \cite{alekseyenko2008wad}
extended the model and give applications in genetics. In this paper we extend the inference to handle
data missing at random. When trait data are gathered, traits are thinned in a way that depends on both the trait and missing-data content.
\citet{nicholls2006dat} treat missing records as
absent traits. Hittite has
12\% missing trait records. Its age is poorly predicted in their
cross-validation. Our prediction is consistent with the historical
record. \citet{nicholls2006dat} dropped seven 
languages with too much missing data. We fit all twenty four
languages in the lexical data of \citet{ringedata}. In order to model spatial-temporal rate heterogeneity we add
a catastrophe process to the model. When a
language passes through a catastrophe, many traits change at
the same time. We fit the full model in a Bayesian setting, via MCMC.
 We validate our fit using Bayes factors to test known age constraints. We reject three of thirty historically attested constraints. Our main result is a unimodel posterior distribution for the age of Proto-indo-European centered at 8400 years BP with 95\% HPD equal 7100-9800 years BP.
\end{abstract}

\begin{newsec} 
The Indo-European languages descend from a common ancestor called
Proto-Indo-European. Lexical data show the patterns
of relatedness among Indo-European languages. These data are
``cognacy classes'': a pair of words in the same class descend, through a process of sound change,
from a common ancestor. For example, English \emph{sea} and German
\emph{See} are cognate to one another, but not to the French
\emph{mer}.
\cite{gray2003ltd} coded data of this kind in a matrix in which rows correspond to languages and columns to distinct cognacy classes, and entries are zero or one as the language possessed or lacked a term in the column class. They
analysed these data using phylogenetic algorithms similar to those used for
genetic data. Our analysis has the same objectives, but we fit a
model designed for lexical trait data. We work
with data compiled by \citet{ringedata}, recording the
distribution of some 872 distinct cognacy classes in twenty four modern and ancient
Indo-European languages. In section \ref{results}, we give estimates
for the unknown topology and branching times of the phylogeny of the
core vocabulary of these languages.

\citet{nicholls2006dat} analyse the same data using a closely related stochastic Dollo-model for binary trait evolution. However, those authors were unable to deal with missing trait
records. Missing data arise when we are unable to answer the question ``does language X possess a cognate in cognacy class Y?''. \citet{nicholls2006dat} dropped seven languages which had many missing entries, and treated missing trait records in the remainder as absent traits. This is
unsatisfactory. However, it is not straightforward to give a
model-based integration of missing data for the trait evolution
model of \citet{nicholls2006dat}. In this paper we integrate the
missing trait data, and this technical advance allows us to fit all
twenty four of the languages in the original data. The binary trait
model of \citet{nicholls2006dat}, has been extended by
\cite{alekseyenko2008wad} to mutliple-level traits, and is finding applications
in biology. A proper treatment of missing data will be of use in
other applications.

We are specifically interested in phylogenetic dating. Because we
are working with lexical, and not syntactic data, it is the age of
the branching of the core vocabulary of Proto-Indo-European that we
estimate. This is a controversial matter. Workers in historical
linguistics have evidence from linguistic paleontology that the most
recent common ancestor of all known Indo-European languages branched
no earlier than about 6000--6500 years Before the Present (BP)
\citep{mallory1989sie}. For a recent review of the argument from
linguistic paleontology, and a criticism of phylogenetic dating, see
\cite{garrett2006cfi} and \cite{mcmahon2005lcn}. An alternative hypothesis
suggests that the spread began around 8500 BP when the Anatolians
mastered farming \citep{renfrew1987aal} in the early neolithic.
Recent efforts to apply quantitative phylogenetic methods to dating
Proto-Indo-European give a time depth of 8000 to 9500 years BP
\citep{nicholls2006dat, gray2003ltd}, supporting the link to
farming. In this work we obtain a unimodal posterior distribution for the age of Proto-Indo-European centered at 8400 years BP with 95\% HPD equal 7100-9800 years BP

One point of view is that the argument from linguistic paleontology
is correct, and the phylogenetic dates are incorrect, due to rate
heterogeneity, or some other model failing. In this respect we
advance the search started in \citet{nicholls2006dat} for a model
mispecification which might explain the 20\% difference the central age
estimates from our fitting, and the status quo. The difference is large enough that we are hopeful of
finding a single coherent error, if any such error exists.
Accounting now for missing data, we are able to publish a much wider
cross-validation test (up from 10 to 30 tested calibrations). Also,
we fit a model for explicit rate heterogeneity in time and space. In
view of the spatial-temporal homogeneity we find here and in other
data, the case for the earlier date seems now fairly strong. The
most probable alternative seems to be a step change in the rate of
lexical diversification acting in a coordinated fashion across the
Indo-European territory some 3000 to 5000 years ago.

In Sections \ref{data} and \ref{prior} we describe the data and specify
a subjective prior for the phylogeny of vocabularies. In sections \ref{evolution} and \ref{likelihood},
we give a generative model for the data and the corresponding likelihood.
We include, in Section \ref{likelihood}, a recursive algorithm
which makes the sum over missing data tractable. In sections \ref{posterior} and
\ref{mcmc} we give the posterior distribution on tree and parameter
space, and briefly describe our MCMC sampler. In section \ref{validation},
we discuss likely model mispecification scenarios, test for
robustness by fitting synthetic data simulated under such
conditions, and cross-validate our predictions. We fit the model to
a data set of Indo-European languages in section \ref{results}. This
paper has a supplement giving an analysis of a second data set,
collected by \citet{dyendata}.

Phylogenetic methods have been used to make inference for tree
\citep{ringedata} and network structure \citep{mcmahon2005lcn} in Historical
Linguistics. \citet{warnow2004sml} write down a more realistic model for
the diversification of vocabulary, accounting for word-borrowing between
languages (so that their history need not be tree-like)
but there is to date no fitting. 

\end{newsec}

\begin{newsec} 
\section{Description of the data}\label{data}

The data group words from 328 meaning categories and 24
Indo-European languages into 872 homology
classes. The data were collected and coded by \citet{ringedata}.
Meaning categories cover the ``core'' vocabulary and are assumed
relevant to all languages in the study. Two words of closely similar
meaning, descended from a common ancestor but subject to variation
in phonology, are {\it cognate} terms. A {\it cognacy class} is a
homology class of words all belonging to a single meaning category.
For example, for the meaning ``head", the Italian \emph{testa} and
the French \emph{t\^ ete} belong to the same cognacy class, while
the English \emph{head} and the Swedish \emph{huvud} belong to
another cognacy class. An element of a cognacy class is thus a word
in a particular language, a sound-meaning pair. These
elements are called {\it cognates}. The vocabulary of a single language is
represented as a set of distinct cognates. In our analysis a cognate has just two properties: its language and its cognacy class. If there are
$N$ distinct cognacy classes in data for $L$ languages, then the
$a$'th class $M_a\subseteq \{1,2,...,L\}$ is a list of the indices
of languages which possess a cognate in that class.
The data are coded as a binary matrix $D$. A row corresponds to
a language and a column to a cognacy class, so that $D_{i,a}=1$ if
the $a$'th cognacy class has an instance in the $i$'th language, and
$D_{i,a}=0$ otherwise. See Table \ref{sea} for an example. This
coding allows a language to have several words for one meaning (such
as Old High German \emph{stirbit} and \emph{touwit} for ''he dies", an instance of
polymorphism), or no word at all. Missing matrix elements arise
because the reconstructed vocabularies of some ancient languages are
incomplete. If we are unable to answer the question ``does language
$i$ possess a cognate in cognacy class $a$'' then we set
$D_{i,a}=?$.

We need notation for both matrix and set representations with
missing data. Denote by $B_a$ column $a$ of $L\times N$ matrix $B$.
For $a=1,2,...,N$ let ${\cal D}_a$ be the set of all
column vectors $d^*$ allowed by the data $D_a$ in column $a$ of $D$,
\begin{equation*}
{\cal D}_a=\left\{d^*\in\{0,1\}^L: D_{i,a}\in\{0,1\}\Rightarrow
d^*_a=D_{i,a},\ i=1,2,...,L\right\}.
\end{equation*}
For $d^*\in{\cal D}_a$ let $m(d^*)=\{i: d^*_i=1\}$. Denote by
$\Omega_a$ the set of cognacy classes consistent with the data
$D_a$, so that
\[
\Omega_a=\{\omega\subseteq \{ 1,2,\ldots,N\} : \omega=m(d^*), d^*\in{\cal D}_a\}.
\]
The data $D$ are then equivalently
$\Omega=(\Omega_1,\Omega_2,...,\Omega_N)$. The $\Omega_a$-notation 
generalizes the $M_a$-notation to handle missing data. It is illustrated in Table \ref{sea}.

\begin{table}
$\begin{array}{c@{\hspace{.3in}}c}
 \begin{tabular}{|>{\bfseries}c|>{\itshape}c|}
\hline
Old English & stierf\th \\\hline
Old High German & stirbit, touwit \\\hline
Avestan & miriiete \\\hline
Old Church Slavonic & um\u\i ret\u u \\\hline
Latin & moritur\\\hline
Oscan & ?\\\hline
\end{tabular}&
\begin{tabular}{|>{\bfseries}c|ccc|}
\hline
Old English & 1&0&0\\\hline
Old High German &1&1&0 \\\hline
Avestan & 0&0&1\\\hline
Old Church Slavonic & 0&0&1 \\\hline
Latin & 0&0&1\\\hline
Oscan & ?&?&?\\\hline
 \end{tabular}\\\\
(a)&(b)\\\\
\end{array}$
\\
\caption{\label{sea} An example of data coding: (a), the word
``he dies'' in six Indo-European languages; (b), the coding of this data
as a binary matrix with ?'s for missing data. In the notation of Section \ref{data}, the first column in (b) is a cognacy class  $M_{\mbox{dies}}\in \Omega_{\mbox{dies}}$ with
$\Omega_{\mbox{dies}}=\{ \{\mbox{Old English, Old High German}\},\
\{\mbox{Old English, Old High German, Oscan}\}\}$}
\end{table}

\citet{ringedata} list twenty four mostly ancient languages. For
eleven of these languages (Latin, Modern Latvian, Old Norse...), all
the matrix entries are recorded. For the rest, the proportion of
missing entries varies between 1\% (for Old Irish) and 91\% (for
Lycian, an ancient language of Anatolia). Note that data are usually
missing in small blocks corresponding to the cognacy classes for a
given meaning category, as in Table \ref{sea}. We do not model this
aspect of the missing data. This is related to the
model-error \citet{nicholls2006dat} call `the empty-field
approximation', under which cognacy classes in the same meaning
category are assumed to evolve independently.

For these Indo-European phylogenies, the topology of some subtrees
are known from historical records. We have lower and upper bounds
for the age of the root node of some subtrees. For example, the
Slavic languages are known to form a subtree, and the most recent
common ancestor of the Slavic languages in the data is known to be
at least 1300 years old. In this way internal nodes of the phylogeny
are constrained. We have age bounds for leaf nodes as
well, since we are told when the vocabulary of the ancient languages
in these data was in use. We combine these {\it calibration} constraints
with the cognacy class data in section~\ref{prior}. Jumping ahead to our results,
Figure~\ref{ie8out4post} is a sample from the posterior distribution
we find for phylogenies. Calibration constraints are represented
by the black bars across nodes in this tree.

\end{newsec}
\begin{newsec} 
\section{Models}\label{model}

We specify a subjective prior for phylogenies, representing a state
of knowledge of interest to us. The model we give for the diversification of vocabulary
in Section~\ref{evolution} is, in contrast, generative.
We model the diversification of vocabulary as a branching process of
sets of cognates, evolving on the phylogeny. Leaves are labeled by languages and a
branch represents the ancestral lineage of a vocabulary.
\begin{newsec}
\subsection{Prior distribution on trees}
\label{prior} The material in this subsection follows
\citet{nicholls2006dat}. We consider a rooted binary tree $g$ with
$2L$ nodes: $L$ leaves, $L-2$ internal nodes, a root node $r=2L-1$
and an Adam node $A=2L$, which is linked to $r$ by a edge of
infinite length. Each node $i=1,2,...,2L$ is assigned an age $t_i$
and $t=(t_1,t_2,...,t_A)$; the units of age are years before the
present; for the Adam node, $t_A=+\infty$. The edge between parent
node $j$ and child node $i$ is a directed branch $\langle
i,j\rangle$ of the phylogeny, with the orderings $i<j$ and
$t_i<t_j$. Let $E$ be the set of all edges, including the edge
$\langle r,A\rangle$, let $V$ be the set of all nodes and
$V_L=\{1,2,...,L\}$ the set of all leaf nodes. Our base tree space
$\Gamma$ is the set of all rooted directed binary trees
$g=(E,V,t)$, with distinguishable leaves and, for $i=1,2,...,2L$,
node ages $t_i>0$ assigned so that the directed path from a leaf
$i\in V_L$ to node $A$ passes through nodes of strictly increasing
age. With this numbering convention, $(E,V)$ is called the ordered
history of a rooted directed binary tree.

We allow for $C$ calibration constraints on tree-topology and
selected node ages. These are described at the end of Section~\ref{data}.
Each such constraint $c$ allows trees in just some
subspace $\Gamma^{(c)}\subset\Gamma$ of tree space. We add to these
constraints an upper bound on the root time at some age $T>0$, and
set $\Gamma^{(0)}=\{(E,V,t)\in \Gamma: t_r\le T\}$. The space of
calibrated phylogenies with catastrophes is then
\[
\Gamma^{C}=\bigcap_{c=0}^C \Gamma^{(c)}.
\]

The root age is a sensitive statistic in this inference. A prior
distribution on trees, with the property that the marginal
distribution of the root age is uniform over a fixed prior range
$t_L\le t_r\le T$, is of interest. For node $i\in V$, let $t^+_i=\sup_{g\in
\Gamma^{C}}t_i$ and $t^-_i=\inf_{g\in \Gamma^{C}}t_i$ be the
greatest and least admissible ages for node $i$, and let $S=\{i\in V: t^+_i=T\}$,
so that $S$ is the set of nodes having ages not bounded above by a
calibration (there are 12 such nodes in Figure~(\ref{ie8out4post})).
\citet{nicholls2006dat} show that, before calibration
(when $C=0$) and for tree spaces in which the leaves
have fixed equal ages, the prior probability distribution with
density
\[f_G(g|T)\propto \prod_{i\in S}(t_r-t^-_i)^{-1}\]
gives a marginal density for $t_r$ which is exactly uniform in
$t_L<t_r<T$. \citet{nicholls2006dat} do not comment on the
distribution determined by $f_G$ over tree topologies. The prior
$f_G$ has in this case ($C=0$) a uniform marginal
distribution on ordered histories exactly equal to the corresponding distribution for
the Yule model. This is not uniform, but favors
balanced leaf-labeled topologies. For $C>0$, and on catastrophe-free tree spaces in
which leaf ages are constrained by prior knowledge to lie in an
interval only, \citet{nicholls2006dat} argue from simulation studies
that this prior gives a reasonably flat marginal distribution for
$t_r$ if in addition $T\gg \max_{i\in V\setminus S} t_i^+$ (the
greatest upper bound among the calibration constraints is not too
close to the upper bound on the root age). We give a sample from the prior in the Supplementary Material; the prior does not represent any reasonable prior belief before about 4500 BP, but this is immaterial as the likelihood rules these values out (an instance of an application of the principle of sufficient reason).

\end{newsec}

\begin{newsec}
\subsection{Diversification of cognacy classes}\label{evolution}

In this subsection we extend the stochastic Dollo model of
\citet{nicholls2006dat} to incorporate rate heterogeneity
in time and space, via a catastrophe process.

Consider the evolution of cognates down the ancestral lineage of
the vocabulary of a single language.
An example is given in Figure~\ref{extree4b}. A new cognacy class is
born when its first cognate is born. This new word is not cognate with other 
words in the modeled process. Loan words from outside the core meaning 
categories of any
language in the study, or from a language outside the study, may be
good for word birth without violating this condition. A cognate dies
in a particular vocabulary when it ceases to be used for the meaning
common to its cognacy class: its meaning may change, or it may fall
out of use.

Cognates evolving in a single language ($ie$ down a single branch of
a language phylogeny) are born independently at rate $\lambda$, die
independently at {\it per capita} rate $\mu$, and are subject to
point-like catastrophes, which they encounter at rate $\rho$ along a
branch. At a catastrophe, each cognate dies independently with
probability $\kappa$, and a Poisson number of cognates with mean
$\nu$ are born. At a branching event of the phylogeny,
the set of cognates representing the branching
vocabulary is copied into each of the daughter languages.
See Figure~\ref{extree4b}.

\begin{figure}[t]
\centering
\makebox{\includegraphics[width=11cm]{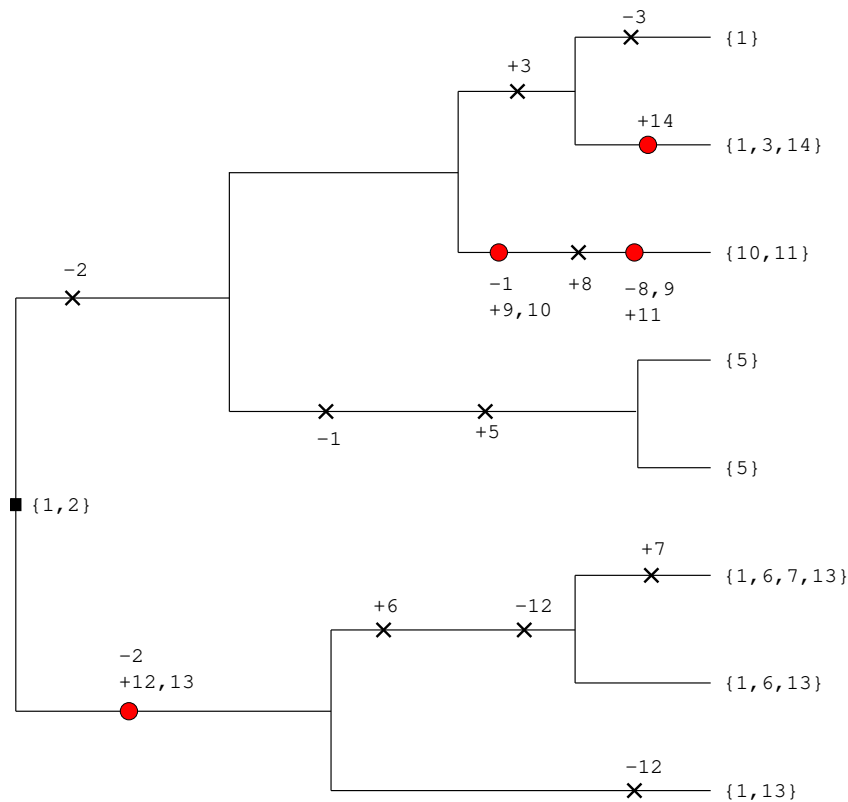}}
\caption{\label{extree4b} Description of the model: births and
deaths of traits are marked. The dots correspond to catastrophes, at
which multiple births and deaths may occur simultaneously. The
cognate sets this generates at leaves are shown on the right. Calendar
time flows from left to right and the age variables $t_i$ in the
text increase from right to left.}
\end{figure}


The process we have described is not reversible, and this greatly
complicates the analysis. It seems acceptable, from a data-modeling perspective,
to impose the condition $\nu=\kappa\lambda/\mu$, which is
necessary and sufficient for reversibility (see Supplementary Material for a proof). 
Under this condition, adding a
catastrophe to an edge is equivalent to lengthening that edge by
$T_C(\kappa,\mu)=-\log(1-\kappa)/\mu$ years. This follows because the number of
cognates generated by the anagenic part of the process in an interval
of length $T_C$
is Poisson distributed with mean $\frac{\lambda}{\mu}(1-e^{-\mu
T_C})$ equal to $\kappa\lambda/\mu$, and the probability that a
cognate entering an interval of length $T_C$ dies during that
interval is $1-e^{-\mu T_C}$, which equals $\kappa$.

Because a catastrophe simply extends its edge by a block of virtual
time, the likelihood depends only on the number of catastrophes on
an edge, and not their location in time. Let $k_i$ be the number of
catastrophes on edge $\langle i,j\rangle$, and
$k=(k_1,\ldots,k_{2L-2})$ be the catastrophe state vector. We record
no catastrophes on the $\langle R,A\rangle$ edge (its length is
already infinite). The tree $g=(V,E,t,k)$ is specified by its
topology, node ages and catastrophe state. Calibrated tree space
extended for catastrophes is
$$\Gamma^C_K=\{(V,E,t,k): (V,E,t)\in \Gamma^C, k\in \mathbb{N}_0^{2L-2}\}.$$
We drop the catastrophe process from the calculation
in Section~\ref{likelihood}. It is straightforward to restore it,
and we do this in the expression for the posterior distribution in Section \ref{posterior}.

\end{newsec}

\begin{newsec}
\subsection{The registration process}\label{observation}


Let ${\bf D}^*$ denote a
notional {\it full} random binary data matrix,
representing the outcome
of the diversification process of Section~\ref{evolution}. The number of columns in ${\bf D}^*$
is random, and equal to $N^*$.
For the realization depicted in Figure~\ref{extree4b}, ${\bf D^*}=D^*$ with $D^*$ displayed in
Figure~\ref{Dstartable}.
\begin{figure}
\begin{verbatim}
[     D*     ]  [     I*     ]  [   D~       ]  [      I     ]  [      D     ]
10000000000000  11111111111111  10000000000000  1   111  1  11  1   000  0  00
10100000000001  01011111111111  ?0?00000000001  0   111  1  11  ?   000  0  01
00000000011000  10111001110110  0?000??001?00?  1   100  1  10  0   0??  1  0?
00001000000000  11111111111111  00001000000000  1   111  1  11  0   100  0  00
00001000000000  11110111111111  0000?000000000  1   011  1  11  0   ?00  0  00
10000110000010  10111111111111  1?000110000010  1   111  1  11  1   011  0  10
10000100000010  11111111111111  10000100000010  1   111  1  11  1   010  0  10
10000000000010  11111111111100  100000000000??  1   111  1  00  1   000  0  ??
\end{verbatim}
  \caption{Registration of the vocabulary realized in
  Figure~\protect\ref{extree4b} supposing a masking matrix $I^*$, as above.
  $D^*$ is the unobserved full data with a column for each cognacy class and a
  row for each language of Figure~\protect\ref{extree4b} (thus cognate 1 is present
  in rows 1,2,6,7 and 8);
  zeros in the masking matrix $I^*$ indicate missing matrix
  elements. Some cognacy classes are then thinned
  (in this example, the
  registration rule keeps cognacy classes with instances displayed in
  one or more language) to
  give the registered data $D$.
  }
\label{Dstartable}
\end{figure}
Note that $D^*$ has a column for each cognacy class present at the root node,
or born below it, whether or not the cognacy class has any cognates represented
in any leaf languages.

We call the random mapping of the unknown full representation ${\bf D}^*$
to the registered data $\bf D$, which is a random matrix with $N$ columns, the {\it registration} process.
There are two stages to this process: the masking of matrix elements of the fully
realised data $D^*$ with ?'s to form an intermediate data matrix
$\tilde D$, and the selection of columns from $\tilde D$
to determine the realised data $D$.

Let ${\bf I}^*$ be a random $L\times N^*$ indicator matrix of
independent Bernoulli random variables for observed elements. The zeros of $I^*$ mark
matrix entries in $D^*$ which will be unobservable.
For $i=1,2,...,L$ and given $a=1,2,...,N^*$, let $\xi_i=\Pr({\bf I}^*_{i,a}=1)$.
The probability $\xi_i$ that
we can answer the question, ``does language
$i$ possess a word in the $a$'th cognacy class?'',
is assumed to be a function of the language index $i$ only. If we get
an answer, it is assumed correct.
Let $\xi=(\xi_1,\ldots,\xi_L)$ and denote by $I^*$ a
realisation of ${\bf I}^*$.
Denote by $\tilde D=\tilde D({D}^*,{I}^*)$ the masked version
of the full random data matrix: if $I^*_{i,a}=1$ then
$\tilde D_{i,a}=D^*_{i,a}$ and if $I^*_{i,a}=0$ then $\tilde D_{i,a}=?$.

Matrix columns may be missing too, so that $N\le N^*$.
We get missing columns even when there are no missing data. The matrix $\tilde D$ in
Figure~\ref{Dstartable} includes some columns with only zeros and question marks,
corresponding to cognacy classes which existed in the past but are
observed in none of the leaf-languages from which the data were
compiled. These cognacy classes are not included in the registered
data.
Denote by $R$ the registration rule ${\bf D}=R(\tilde {\bf D})$
mapping the full data to registered data.

Rule $R$ may thin additional column types.
Let $Y$ and $Q$ be functions of the columns of $D^*$ and $I^*$
counting the visible 1's and ?'s respectively,
\begin{eqnarray*}
  Y({\bf D}^*_a,{\bf I}^*_a) &=& \sum_{i=1}^L {\bf I}^*_{i,a}{\bf D}^*_{i,a}, \\
  Q({\bf I}^*_a) &=& \sum_{i=1}^L (1-{\bf I}^*_{i,a}).
\end{eqnarray*}
Given $a=1,2,...,N^*$, let $Y_a=Y({\bf D}^*_a,{\bf I}^*_a)$ and $Q_a=Q({\bf I}^*_a)$.

In Appendix~\ref{nounique} we give an efficient algorithm for computing
the likelihood for rules formed by compounding the following elementary thinning operations:
\begin{enumerate}
\item[(1)] $R_1(\tilde D)=(\tilde D_a: Y_a>0)$ (discard classes with no instances at the leaves);
\item[(2)] $R_2(\tilde D)=(\tilde D_a: Y_a>1)$ (discard classes - singletons - observed at a single leaf);
\item[(3)] $R_3(\tilde D)=(\tilde D_a: Y_a<L)$ (discard classes which are observed at all leaves);
\item[(4)] $R_4(\tilde D)=(\tilde D_a: Y_a<L-1)$ (discard classes which are observed at all leaves
                  or at all leaves but
one);
\item[(5)] $R_5(\tilde D)=(\tilde D_a: Y_a+Q_a<L)$ (discard classes which are potentially present at all
leaves);
\item[(6)] $R_6(\tilde D)=(\tilde D_a: Y_a+Q_a<L-1)$ (discard classes which are potentially present at all leaves or at all leaves but
one).
\end{enumerate}
We assume the chosen rule includes Condition (1).
The rule $D=R(\tilde D)$ with $$R(\tilde D)=R_4\circ R_2(\tilde D)$$
collects ``parsimony informative'' cognacy classes.
\citet{ronquist2005mm} give the likelihood for the finite-sites trait
evolution model of \cite{lewis2001lae} for registration rules like (1-6).
In the example in Figure~\ref{Dstartable}, and in
Sections~\ref{reconstruction}~and~\ref{results}) we fit data
registered with $R(\tilde D)=R_1(\tilde D)$.

The selection of columns is something we have in general no control over:
the column selection rule simply describes what happened at registration.
Results in the Supplementary Material for the
\citet{dyendata} data use the rule $R(\tilde D)=R_2(\tilde D)$, since
singleton columns were not included in that data.
However, certain column types may include data which is hard to model well, and so we
may choose to make further thinning using the other rules. Recursions for
the other rules are given in an Appendix. 

Column indices $a=1,2,...,N^*$ are exchangeable. It is convenient to
renumber the columns of $D^*$, $I^*$ and $\tilde D$ after
registration, so that $\tilde D_a=D_a$ and $I^*_a=I_a$ for $a=1,2,...,N$. The
information needed to evaluate $Y_a$ and $Q_a$ is available in the
column $D_a$ and set $\Omega_a$ representations. We write
$Y(D_a)=Y(\Omega_a)=Y_a$ and $Q(D_a)=Q(\Omega_a)=Q_a$.

\end{newsec}
\begin{newsec} 
\subsection{Point process of births for registered cognacy classes}

Fix a catastrophe-free phylogeny $g\in \Gamma^C_K$, with
$k=(0,0,...,0)$, and let an edge $\langle i,j\rangle$ and a time
$\tau\in [t_i,t_j)$ be given. Denote by $[g]$ the set of all points
$(\tau,i)$ on the phylogeny, including points $(\tau,r)$ with
$\tau\ge t_r$ in the edge $\langle r,A\rangle$. The locations
$z_D=\{z_1,z_2,...,z_N\}$ of the birth events of the $N$ {\it registered}
cognacy classes are a realization of an inhomogeneous Poisson point
process $Z_D$ on $[g]$. Let $Z\in [g]$ be the birth location of a
generic (and possibly unregistered) cognacy class $M\subseteq \{1,2,...,L\}$, corresponding to a
column of $\tilde {\bf D}$ with $Y$ observed $1$'s and $Q$ $?$'s, and
let ${\cal E}_Z$ be the event that this class generates a column of
the registered data. For the $a$'th cognacy class $M_a$, born at
$Z_a$, this event is ${\cal E}_{Z_a}=\{{\bf D}_a=R(\tilde {\bf
D}_a)\}$ since $R(\tilde{\bf D}_a)$ is empty for $\tilde{\bf D}_a$ a
column dropped at registration.

Cognacy classes are born at constant rate on the branches of $g$,
but are thinned by registration. However, conditional on the birth
location $Z_a=z_a$, our modeling assumes that the outcome ${\cal
E}_{Z_a}$ for the $a$'th cognacy class is decided independently of
all events in all other cognacy classes. The point process $Z_D$ of
birth locations of registered cognacy classes has intensity $$\tilde
\lambda(z)=\lambda\Pr({\cal E}_{Z}|g,\mu,\lambda,\xi,Z=z)$$ at $z\in
[g]$ and probability density
\[
f_{Z_D}(z_D)=\frac{1}{N!}e^{-\Lambda([g])}\prod_{a=1}^N \tilde\lambda(z_a)
\]
with respect to the element of volume $dz_D=dz_1 dz_2...dz_N$ in $[g]^N$, where
\begin{eqnarray*}
  \Lambda([g])&=&\int_{[g]}\tilde\lambda(z)dz\\
  &=&\sum_{\langle i,j\rangle\in E}\int_{t_i}^{t_j} \tilde \lambda((\tau,i)) d\tau.
\end{eqnarray*}
The number $N$ of registered cognacy classes is $N\sim\mbox{Poisson}(\Lambda([g]))$.
%
%
\end{newsec}

\end{newsec}

\begin{newsec} 
\section{Likelihood calculations}\label{likelihood}

We give the likelihood for $g$, $\mu$, $\lambda$, $\kappa$, $\rho$
and $\xi$ given the data, along with an efficient algorithm to
compute the sum over all missing data. The catastrophe process is
left out, and reincorporated in the next section.

Since we only ever see registered data, the likelihood for
$g,\lambda,\mu$ and $\xi$ is the probability
$P[{\bf D}=D|g,\mu,\lambda,\xi,{\bf D}=R(\tilde{\bf D})]$, to get data $D$ given
the parameters and conditional on
the data having passed registration. We restore the
birth locations (and so omit $\lambda$ from the conditioning),
and factorize using the joint independence of $D_a, a=1,2,...,N$
under the given conditions:
\begin{eqnarray*}
  P[{\bf D}\!=\!D|g,\mu,\lambda,\xi,{\bf D}\!=\!R(\tilde{\bf D})] &\!=\!&
  \int f_{Z_D}(z_D)P[{\bf D}\!=\!D|g,\mu,\xi,Z_D\!=\!z_D,{\bf D}\!=\!R(\tilde{\bf D})]\, dz_D, \\
    &\!=\!& \frac{e^{-\Lambda([g])}}{N!}\prod_{a=1}^N \lambda
                \int_{[g]} \!\!P[{\cal E}_{Z_a}|g,\mu,\xi,Z_a\!=\!z_a]\
                \! P[{\bf D}_a\!=\!D_a|g,\mu,\xi,Z_a\!=\!z_a,{\cal E}_{Z_a}]\, dz_a,\\
    &\!=\!& \frac{e^{-\Lambda([g])}}{N!}\prod_{a=1}^N \lambda
                \int_{[g]} \!\! P[{\bf D}_a\!=\!D_a|g,\mu,\xi,Z_a\!=\!z_a]\, dz_a.
\end{eqnarray*}
The last line follows because $P[{\cal E}_{Z_a}|{\bf
D}_a=D_a,...,Z_a=z_a]=1$ for $D_a$ a column of registered data: if
the outcome of the birth at $z_a$ was the registerable data $D_a$
then the event ${\cal E}_{Z_a}$ certainly occurs. The likelihood
depends on the awkward condition ${\bf D}=R(\tilde{\bf D})$ through
the mean number $\lambda([g])$ of registered cognacy classes only,
while $P[{\bf D}_a=D_a|g,\mu,\xi,Z_a=z_a]$ is the probability to
realise the data vector $D_a$ in the unconditioned
diversification/missing element process. The calculation has so far
extended \citet{nicholls2006dat} to give the likelihood for a
greater variety of column thinning rules. We now add the missing
element component of the registration process.

We sum over possible values of the missing matrix elements in the registered data.
Since $P[{\bf D}_a=D_a|g,\mu,\lambda,\xi,Z_a=z_a]$ is not
conditioned on the requirement that the column $D_a$ gets registered,
the entries of the corresponding column $I_a$ are determined by the unconditioned
Bernoulli process, and we have
\begin{eqnarray*}
P[{\bf D}_a=D_a|g,\mu,\xi,Z_a=z_a] &=& \sum_{d^*\in {\cal D}_a} P[{\bf I}^*_a,{\bf D}^*_a=d^*|g,\mu,\xi,Z_a=z_a] \\
&=& \prod_{i=1}^L \xi_i^{I_{i,a}}(1-\xi_i)^{1-I_{i,a}} \sum_{d^*\in {\cal D}_a} P[{\bf D}^*_a=d^*|g,\mu,\xi,Z_a=z_a].
\end{eqnarray*}
The likelihood is
\begin{equation}
\label{eq:lkd}
  P[{\bf D}=D|g,\mu,\lambda,\xi,{\bf D}=R(\tilde{\bf D})] =
                \frac{e^{-\Lambda([g])}}{N!}\prod_{a=1}^N (\prod_{i=1}^L \xi_i^{I_{i,a}}(1-\xi_i)^{1-I_{i,a}} ) \lambda
                \int_{[g]} \sum_{\omega\in \Omega_a}P[M=\omega|g,\mu,\xi,Z=z_a] dz_a,
\end{equation}
where we have switched now from summing $d^*\in{\cal D}_a$
to the equivalent set representation $\omega\in\Omega_a$.

For the two integrated quantities in
Equation~(\ref{eq:lkd}) we have tractable recursive formulae.
We are using a pruning procedure akin to \citet{felsenstein1981ip}.
We begin with $\Lambda([g])$.

We assume the registration rule includes at least Condition (1). It follows that
a cognacy class born at $Z=(\tau,i)$ in $[g]$ must survive down to the node below,
at $Z=(t_i,i)$, in order to be registered, and so
$$P[{\cal E}_{Z}|Z=(\tau,i),g,\mu,\xi]=P[{\cal E}_{Z}|Z=(t_i,i),g,\mu,\xi]e^{-\mu(\tau-t_i)}.$$
We can substitute this into the expression for $\Lambda([g])$, and integrate, to get
\begin{equation}
 \Lambda([g])=\frac{\lambda}{\mu}\sum_{\langle i,j\rangle\in E}
 P[{\cal E}_{Z}|Z=(t_i,i),g,\mu,\xi]\left(1-e^{-\mu(t_j-t_i)}\right). \label{eqA}
\end{equation}
Given a node $i$, let $V_L^{(i)}$ be the set of leaf nodes descended from $i$,
including $i$ itself if $i$ is a leaf. Let $s_i=\mbox{card}(V_L^{(i)})$.
Denote by $u_i^{(n)}=P[Y=n|Z=(t_i,i),g,\mu,\xi]$ and
$v_i^{(n)}=P[Y+Q=n|Z=(t_i,i),g,\mu,\xi]$. We can compute $\Lambda$ for rules
made up of combinations of Condition (1) with any combination of Conditions (2-6),
from $u_i^{(0)}$, $u_i^{(1)}$, $u_i^{(s_i-1)}$, $u_i^{(s_i)}$,
$v_i^{(s_i-1)}$ and $v_i^{(s_i)}$. For example,
\begin{equation}\label{eq:examplereg}
P[{\cal E}_{Z}|Z=(t_i,i),g,\mu,\xi]=\left\{\begin{array}{ll}
                              1-u_i^{(0)} & R=R_1, \\
                              1-u_i^{(0)}-u_i^{(1)} & R=R_2, \\
                              1-u_i^{(0)}-u_i^{(1)}-u_i^{(L-1)}-u_i^{(L)} & R=R_3\circ R_2.\}
                            \end{array}\right.
\end{equation}
Notice that $u_i^{(n)}=0$ unless $s_i\ge n$, so for example
$u_i^{(L)}$ is non-zero at $i$ the root node only.
Since our main application is for data registered under Condition (1),
we give, in the body of this paper,
recursions for $u_i^{(0)}$ and $v_i^{(0)}$ only. See Appendix~\ref{nounique}
for the recursions needed to evaluate the likelihood under
rules involving Conditions (2-6).

For nodes $i$ and $j$, let $\delta_{i,j}=e^{-\mu(t_j-t_i)}$.
Consider a pair of edges $\langle c_1,i\rangle$, $\langle c_2,i\rangle$ in $E$.
\begin{eqnarray}
 u_i^{(0)} & =& \left( (1-\delta_{i,c_1})+\delta_{i,c_1}u_{c_1}^{(0)}\right) \left( (1-\delta_{i,c_2})+\delta_{i,c_2}u_{c_2}^{(0)}\right) \\
 v_i^{(0)} &=& \left(\delta_{i,c_1}v_{c_1}^{(0)} + (1-\delta_{i,c_1})\prod_{j\in V_L^{c_1}}\xi_j\right) \left(\delta_{i,c_2}v_{c_2}^{(0)} + (1-\delta_{i,c_2})\prod_{j\in V_L^{c_2}}\xi_j\right)
\end{eqnarray}
The recursion is evaluated from the leaves $i\in V_L$, at which
\begin{eqnarray}
u_{i}^{(0)}&=&1-\xi_i\\
v_i^{(0)}&=&0\label{eqB}
\end{eqnarray}

We now give the equivalent recursions for
$\lambda \int_{[g]}\sum_{\omega \in\Omega_a}P[M=\omega_a|Z=z_a,g,\mu]\, dz_a$.
Consider the set $m_a=\bigcap_{\omega\in\Omega_a} \omega$ of leaves {\it known} to have a
cognate in the $a$th registered cognacy class.
Let $E_a$ be the set of branches on the path from
the most recent common ancestor of the leaves in $m_a$ up to the Adam-node $A$ above the root.
Cognacy class $a$ must have been born on an edge in $E_a$.
For $a=1,2,...,N$, class $M_a$ is non-empty and we can shift the birth location
to the node below and convert the integral to a sum,
\begin{equation*}
 \lambda \int_{[g]}\sum_{\omega \in\Omega_a}
 P[M=\omega|Z=z_a,g,\mu]\, dz_a = \frac{\lambda}{\mu}\sum_{\langle i,j\rangle \in E_a}
 \sum_{\omega \in \Omega_a} P[M=\omega|Z=(t_i,i),g,\mu](1-\delta_{i,j}).\label{eqC}
\end{equation*}
For each $a=1,2,...,N$ and $\omega\in\Omega_a$, let $\omega^{(i)}= \omega\cap V_L^{(i)}$ and
$$\Omega_a^{(i)}=\{\omega^{(i)}: \omega^{(i)}= \omega\cap V_L^{(i)}, \omega\in\Omega_a\}$$
denote the set of all subsets $\omega^{(i)}$ of the leaves $V_L^{(i)}$ which are cognacy classes consistent with
the data available for those leaves.
Consider two child branches $\langle c_1,i \rangle$ and $\langle c_2,i\rangle$
at node $i$. Since $\Omega_a=\Omega_a^{(c_1)}\times\Omega_a^{(c_2)}$, and
events are independent along the two branches,
\begin{eqnarray*}
\sum_{\omega \in \Omega_a} P[M=\omega|Z=(t_i,i),g,\mu,\xi] &=&
\sum_{\omega^{(c_1)} \in \Omega_a^{(c_1)}} P[M=\omega^{(c_1)}|Z=(t_i,c_1),g,\mu]\nonumber\\
&&\times\hspace*{-8pt}\sum_{\omega^{(c_2)} \in \Omega_a^{(c_2)}} P[M=\omega^{(c_2)}|Z=(t_i,c_2),g,\mu].
\end{eqnarray*}
Having moved the birth event at $(t_i,i)$ to $(t_i,c_1)$ and $(t_i,c_2)$
(off the node and onto its child edges) we now move the birth event at
$(t_i,c)$ to $(t_{c},c)$ (down an edge) as follows:
\begin{eqnarray*}\label{downbranch}
\lefteqn{\sum_{\omega \in \Omega_a^{(c)}}
P[M=\omega|Z=(t_i,c),g,\mu]=} &&  \\
&&\left\lbrace
\begin{array}{ll}
 \delta_{i,c} \times \displaystyle \sum_{\omega \in \Omega_a^{(c)}}
 P[M=\omega|Z=(t_{c},c),g,\mu] & \text{ if } Y(\Omega_a^{(c)})\geq 1 \\
(1-\delta_{i,c}) + \delta_{i,c} \times \displaystyle \sum_{\omega \in \Omega_a^{(c)}} P[M=\omega|Z=(t_{c},c),g,\mu] & \text{ if } Y(\Omega_a^{(c)})=0 \text{ and } Q(\Omega_a^{(c)}) \geq 1 \\
(1-\delta_{i,c}) + \delta_{i,c}v_{c}^{(0)} & \text{ if } Y(\Omega_a^{(c)})+Q(\Omega_a^{(c)})=0 \\
&\quad\text{(\emph{i.e.} } \Omega_a^{(c)}=\{\emptyset\} \text{)}
\end{array}
\right.\nonumber
\end{eqnarray*}
The recursion is evaluated from the leaves. If $c$ is a leaf, then
\begin{equation*}\
\sum_{\omega \in \Omega_a^{(c)}} P[M=\omega|Z=(t_{c},c),g,\mu]=
\begin{cases}
 1 & \text{ if } \Omega_a^{(c)} =\{\{c\},\emptyset\} \text{ or } \{\{c\}\} \text{ (\emph{i.e.} }D_{c,a}\in\{?,1\}\text{)} \\
0  & \text{ if } \Omega_a^{(c)} = \{\emptyset\} \text{ (\emph{i.e.} }D_{c,a}=0\text{)}.
\end{cases}\label{eqD}
\end{equation*}

In order to restore catastrophes to this calculation, and given $g\in \Gamma_K$,
with $k_i$ catastrophes on edge $\langle i,j\rangle \in E$,
replace $t_j-t_i$ with $t_j-t_i+k_iT_C(\kappa,\mu)$ throughout.
\end{newsec}

\begin{newsec} 
\section{Posterior distribution}\label{posterior}

Our prior on the birth rate $\lambda$, death rate $\mu$ and catastrophe rate $\rho$ is $p(\lambda, \mu, \rho)\propto
\frac{1}{\lambda\mu\rho}$ and we take a uniform prior over $[0,1]$
for the death probability at a catastrophe $\kappa$ and each missing
data parameter $\xi_i$.

Substituting using equations
(\ref{eqA})-(\ref{eqD}) into equation (\ref{eq:lkd}) and multiplying
by the prior $f_G(g|T)p(\lambda, \mu, \rho)$, we obtain the posterior distribution
\begin{eqnarray}\label{eq:posterior}
\lefteqn{p(g,\mu,\lambda,\kappa,\rho,\xi|{\bf D}=D)} && \nonumber\\
&= & \frac{1}{N!}\left(\frac{\lambda}{\mu}\right)^N
\exp \left(-\frac{\lambda}{\mu} \sum_{\langle i,j\rangle \in E}P[{\cal E}_{Z}|Z=(t_i,i),g,\mu,\kappa,\xi](1-e^{-\mu(t_j-t_i+k_iT_C)})\right) \nonumber\\
&&\times   \prod_{a=1}^N \left(\sum_{\langle i,j\rangle \in E_a} \sum_{\omega \in \Omega_a} P[M=\omega|Z=(t_i,i),g,\mu](1-e^{-\mu(t_j-t_i+k_iT_C)})\right) \nonumber\\
&&\times
\frac{1}{\mu\lambda\rho}f_G(g|T)\frac{e^{-\rho |g|}(\rho |g|)^{k_T}}{k_T!}\prod_{i=1}^L(1-\xi_i)^{Q_i}\xi_i^{N-Q_i}
\end{eqnarray}
for parameters $\mu,\lambda,\kappa,\rho>0$, $0<\xi_i<1$ and trees
$g\in\Gamma_K^{(C)}$.

The posterior is improper without bounds on $\rho$ since $k_T=0$ is allowed. We place very
conservative bounds on $\rho$. Results are not sensitive to this choice.
We can show that, for 'typical' data sets $D$, and in particular the data
analysed below, the posterior is then proper. Details of the relationship between
cognate classes and calibration constraints play a role in the conditions
for the posterior distribution to be proper.

\end{newsec}

\begin{newsec} 
\section{Markov Chain Monte Carlo}\label{mcmc}

We use Markov Chain Monte Carlo to sample the posterior distribution
and estimate summary statistics.
If the prior on the cognacy class
birth rate parameter $\lambda$ has the conjugate form
$\lambda^u\exp(-v\lambda)$ then the
conditional distribution of $\lambda$ in the posterior distribution above
has the form $\lambda^{N+u}\exp(-\lambda (X+v))$.
We took the improper prior $u=-1$, $v=0$ for $\lambda$
and integrated. The MCMC state is then $x=((E,V,t,k),\mu,\kappa,\rho,\xi)$ 
and the target distribution $p(x|D)$ is the density obtained by integrating
the density in Equation~\ref{eq:posterior} over $\lambda$.

The MCMC sampler described in \citet{nicholls2006dat} has state
$x=((E,V,t),\mu)$. We add to the $(E,V,t)$- and $\mu$-updates of
\citet{nicholls2006dat} further MCMC updates
acting on the catastrophe vector $k=(k_1,...,k_{L-2})$, on the catastrophe parameters
$\kappa,\rho$ and on $\xi$, the probability parameter for observable data-matrix elements.
The catastrophe rate parameter $\rho$ is added to time-scaling updates in which
subsets of parameters are simultaneously scaled by
a common random factor $s$: if $\theta$ is a parameter in the scaled subset
having units $[\mbox{time}]^{u}$, then $\theta\rightarrow s^u\theta$.
The probability $0<\xi_i<1$ for an element
of the registered data matrix to be observable is, for many leaf-languages,
close to one, so we update those parameters by scaling $1-\xi_i$.

We incorporate updates adding and deleting catastrophes (the filled dots marked on
the branches of Figure~\ref{extree4b}) plus an
update which moves a catastrophe from an edge to a
parent, child or sibling edge. For the addition and
deletion of catastrophes, we do not need to use reversible jump
Markov Chain Monte Carlo, as the state vector specifies
the numbers, and not the locations, of catastrophes on edges.


We omit the details of these moves but give, as an example, the
update that moves a catastrophe from an edge to a parent, child or
sibling edge. Let $k_T=\sum_{i=1}^{L-2} k_i$ give the total number of catastrophes.
Given a state $x=(g,\mu,\kappa,\rho,\xi)$ with
$g=(V,E,t,k)$, we pick edge $\langle i,j\rangle\in E$ with
probability $k_i/k_T$. Let $E_{\langle i,j\rangle}$ be the set of
edges \emph{neighbouring} edge $\langle i,j\rangle$
(child, sibling and parent edges, but excluding the edge $\langle R,A\rangle$)
and let $q_i=\mbox{card}(E_{\langle i,j\rangle})$.
We have in general $q_i=4$.
However, for $i$ the index of a leaf node,
$q_i=2$ (1 parent, 1 sibling, no children). If $j$ is the
root and $i$ is non-leaf, then $q_i=3$ (1 sibling, 2 children)
and if $j$ is the root and $i$ is a leaf we have $q_i=1$ (a sibling edge).
Choose a neighbouring edge $\langle\tilde i,\tilde j\rangle$
uniformly at random from $E_{\langle i,j\rangle}$
and move one catastrophe from $\langle i,j\rangle$ to
$\langle \tilde i,\tilde j\rangle$. The candidate state is
$x'=((V,E,t,k'),\mu,\kappa,\rho)$, with $k_i'=k_i-1$ and $k_{\tilde
i}'=k_{\tilde i}+1$ and $k'_j=k_j$ for $j\ne i,\tilde i$.
This move is accepted with probability
\begin{equation*}
 \alpha(x'|x)=\min\left(1,\frac{q_i\, k_{\tilde i}'\, p(x'|D)}{q_{\tilde i}\, k_i\, p(x|D)}\right).
\end{equation*}

We assessed convergence with the asymptotic behaviour of the autocorrelation for the parameters $\mu$, $\kappa$, $\rho$ and $t_R$, as suggested by \citet{geyer1992pmc}. This method indicated that we could use runs of about 10 million samples; we also let the MCMC run for 100 million samples and checked that the computed statistics did not vary.

\end{newsec}

\begin{newsec} 
\section{Validation}\label{validation}

We made a number of tests using synthetic data. Fitting the model to
synthetic data simulated according to the likelihood
$P({\bf D}=D|g,\mu,\lambda,\rho,\kappa,\xi,{\bf D}=R(\tilde{\bf D}))$,
(in-model data),
shows us just how informative the data is for catastrophe placement,
as well as making a debug-check on our implementation.
We fit out-of-model data also. These are synthetic data simulated under likely
model-violation scenarios, and are used to identify sources of
systematic bias. We summarise results for synthetic data simulated
using the parameter values we estimate in section \ref{results} on
the data. For in-model data we correctly reconstruct
topology, root age and the number and position of catastrophes.
Further details are given in the Supplementary Material.

\begin{newsec}
\subsection{Model mis-specification}\label{synthetic}

\citet{nicholls2006dat} use out-of-model data representing
un-modeled loan-words (called {\it borrowing}), rate-heterogeneity
in time and space, rate heterogeneity across cognacy classes, and
the empty-field approximation. They discuss also model
mis-specification due to missing data and the incorrect
identification of cognacy classes (in particular, a hazard for
deeply rooted classes to be split). Rate-heterogeneity in time and
space, and missing data are now part of the in-model
analysis.

We synthesized out-of-model data with the borrowing model of \citet{nicholls2006dat},   
in order to see if un-modeled borrowing biased our results. For what \citet{nicholls2006dat} call low
to moderate levels of borrowing, we were able to reconstruct true
parameter values well. See the Supplementary Material for details. With high
levels of borrowing, we under-estimate the root age and
over-estimate rate parameters. However, unidentified loan words in
the registered data do not generate model mispecification unless
they are copies of cognates which occur in the observed meaning
categories of languages ancestral to the leaf-languages. It is not
plausible that unidentified borrowing of this kind is present at
high levels.

We have not repeated the \citet{nicholls2006dat} out-of-model
analysis of rate heterogeneity across cognacy classes.

In a real vocabulary, distinct cognacy classes that share a meaning
category would not evolve independently. Also, a real language might 
be expected to a possess a word in each of the core
meaning categories. This constrains the number of cognacy classes in
each meaning category to be non-zero. Our model allows empty meaning
categories. \citet{nicholls2006dat} find no substantial bias in a
fit to out-of-model data respecting the constraint. 

Our treatment of missing data introduces a new mis-specification. We
modeled matrix elements as missing independently. However, we get
missing data when we do not know the word used in a given language
to cover a given meaning category. Matrix elements are as a
consequence typically missing in blocks corresponding to all the
cognacy classes for the given meaning. Because the ages of poorly
reconstructed languages are well predicted in the cross-validation
study below, we have not looked further at this issue.

\end{newsec}
\begin{newsec} 

\subsection{Prediction tests for calibration}\label{xval}
\label{reconstruction}

In the next section we estimate the age of a tree node (the root). In this sectio nwe test to see if the uncertainties we estimate are reliable.

The calibration data described in Section~(\ref{data})
fix the topologies and root ages of the subtrees
marked with bars in Figure~\ref{ie8out4post}). 
We remove each calibration constraint in turn and use
the data and the remaining constraints to estimate the age of the
constrained node, and the probability for the constrained subtree
topology. The topological constraints were all perfectly
reconstructed. In twenty three of the twenty eight tests the
constrained age interval overlaps the 95\% HPD interval, as shown in
the bottom half of Figure \ref{xvalfig}.

How good or bad are each of these predictions? Looking at the
Hittite prediction in Figure~\ref{xvalfig}, the large prediction
uncertainty only just allows the calibration interval. Is this bad? 
We quantify the goodness-of-fit, for each calibration, using
Bayes factors to replace $p$-values as indices of misfit. For each constraint $c=1,2,...,C$ we compute a Bayes factor measuring the support for the fully constrained model compared to a model with just the $c$'th constraint removed.

For $c=1,2,...,C$ let
\[
\Gamma^{C-c}=\bigcap_{c'=0\atop c'\ne c}^C \Gamma^{(c')}.
\]
denote the enlarged tree space with the $c$'th constraint removed, and let
$\Gamma^{C-c}_K$ be the enlarged tree space, extended to include catastrophes
(as in Section~(\ref{evolution})).
For each constraint $c=1,2,...,C$ we make a model comparison between a common null model
with all the constraints, $H_0: g\in \Gamma^C_K$, and an alternative model
$H_1: g\in \Gamma^{C-c}_K$ with the constraint removed.
The Bayes factor $B_{C,C-c}$ for the model comparison is the ratio of the posterior probabilities for
these models with model prior $P(H_0)=P(H_1)$,
    \begin{eqnarray*}
      B_{-c} &=& \frac{P(D|g\in \Gamma^{C}_K)}{P(D|g\in \Gamma^{C-c}_K)} \\
                                    &=& \frac{P(D|g\in \Gamma^{C}_K\cap \Gamma^{C-c}_K)}{P(D|g\in \Gamma^{C-c}_K)} \\
        &=& \frac{P(g\in \Gamma^{C}_K|D,g\in \Gamma^{C-c}_K)}{P(g\in \Gamma^{C}_K|g\in \Gamma^{C-c}_K)}.
    \end{eqnarray*}
    where the second line follows since $\Gamma^{C}_K\subset \Gamma^{C-c}_K$ and the third
    from the definition of the conditional probabilities.
The numerator $P(g\in \Gamma^{C}_K|D,g\in \Gamma^{C-c}_K)$ is the posterior probability for the
$c$'th constraint to be satisfied given the data and the other constraints. The denominator,
$P(g\in \Gamma^{C}_K|g\in \Gamma^{C}_K)$, is the prior probability for the
$c$'th constraint to be satisfied given the other constraints. We estimate these probabilities
using simulation of the posterior and prior distributions with constraint $c$ removed.
The Bayes factors are estimated with negligible uncertainty and $2\log(B_{C,C-c})$ is
plotted for $c=1,2,...,C$ in the top half of Figure~\ref{xvalfig}.

Strong evidence against a calibration is failure at prediction. Taking
a Bayes factor exceeding 12 (that is, $2\log(B_{C,C-c})\gtrsim 5$ in Figure~\ref{xvalfig})
as strong evidence against the constraint, following \cite{redbook}, we have conflict for three of the thirty
constraints: the ages of Old Irish and Avestan, and for the age of the Balto-Slav clade.
As our analysis in Section~\ref{results} shows, there is a high posterior probability
that a catastrophe event occurred on the branch between Old Irish and Welsh, and 
another between Old Persian and Avestan. The evidence for rate heterogeneity in
rest of the tree is so slight, that when we try to predict these
calibrations we are predicting atypical events.

Our missing-data analysis has helped here.
The calibration interval for the Hittite vocabulary in these data is
3200--3700BP. A reconstruction of the age
of Hittite which ignores missing data predicts 60--2010BP, well
outside of the constraints. The 95\% HPD interval for the age of
Hittite in our model is 430--3250BP, which just overlaps the
constraint. The Bayes factor gives odds less than 3:1 against,
so the evidence against the constraint is `hardly worth mentioning'.

\begin{figure}
\centering
 \makebox{\includegraphics[width=16cm]{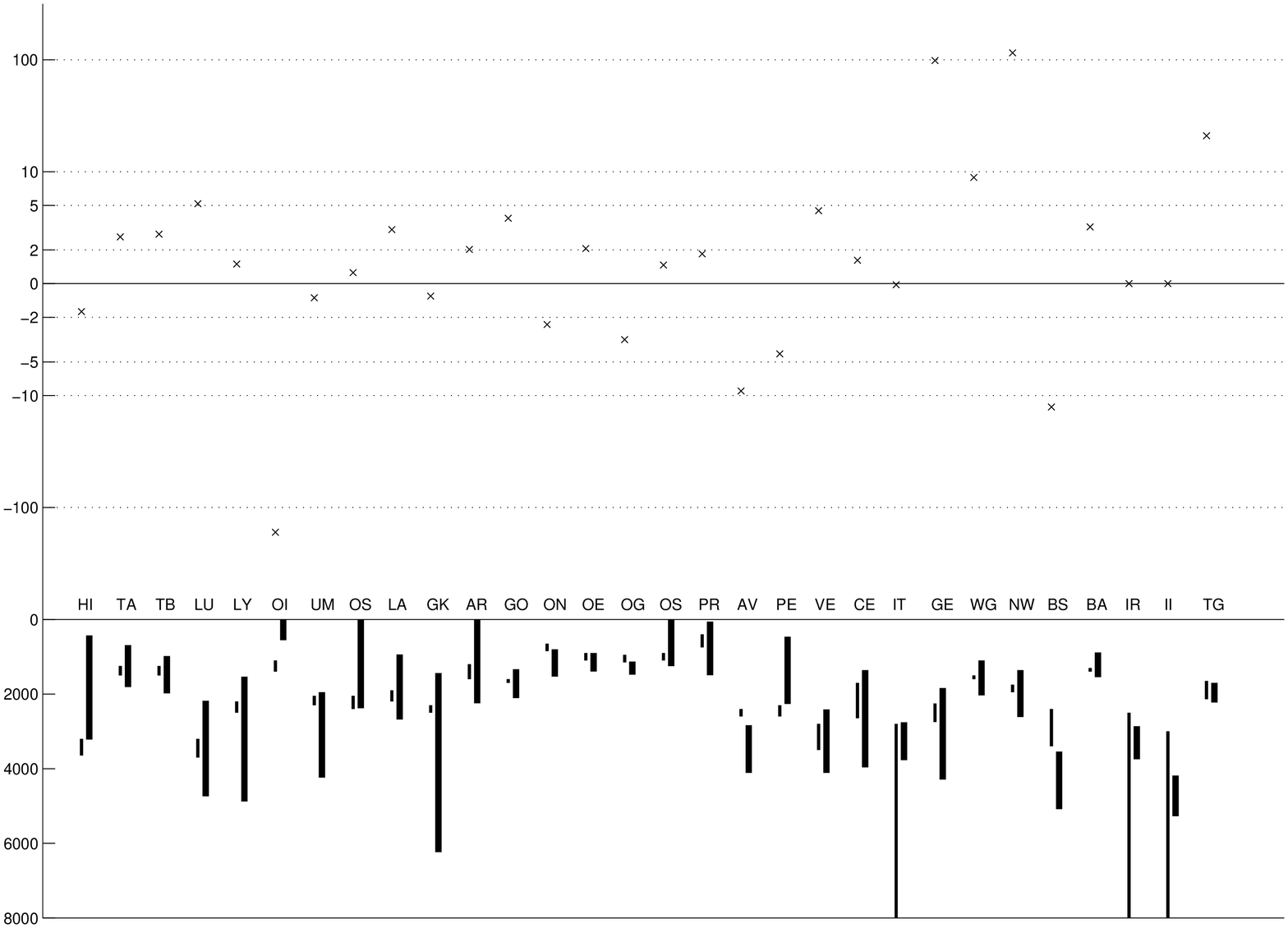}}
\caption{\label{xvalfig}\textit{Reconstruction of known node ages: top, logarithm of Bayes factors $\log(B_{C,C-c})$ for $c=1,2,...,C$;
bottom, thin lines show age constraints for different nodes, thick lines show 95\% posterior HPD interval for the reconstructed dates. The nodes are displayed in the same order as the constraint in Fig. \ref{ie8out4post}}}
\end{figure}

\end{newsec}
\end{newsec}

\begin{newsec} 
\section{Results}\label{results}

In this section we present results for our MCMC simulation of the full posterior,
Equation~(\ref{eq:posterior}). An upper limit $T=16000$ was used in the tree prior of Section~\ref{prior}. Any value for $T$ exceeding around $T=10000$ would lead to the same results.
We show a consensus tree in figure \ref{ringe}. In a tree, an edge corresponds to a split
partitioning the leaves into two sets. A consensus tree
displays just those splits present in at least 50\% of the
posterior sample. Splits which receive less than 95\% support are
labelled. Where no split is present in 50\% of the posterior sample,
the consensus tree is multifurcating. The date shown for a node is
the average posterior date given the existence of the split;
similarly, the number of catastrophes shown on an edge is the
average posterior number of catastrophes on that edge given the
existence of the split, rounded to the nearest integer. Our estimates for the parameters are as follows: $\mu=1.86\cdot10^{-4}\pm1.47\cdot10^{-5}$ deaths/year; $\kappa=0.361\pm 0.055$; $\rho=6.4\cdot10^{-5}\pm3.3\cdot10^{-5}$ catastrophes/year (corresponding to large but rare catastrophes: about 1 catastrophe every 15,000 years, or an average of 3.4 on the tree, with each catastrophe corresponding to 2400 years of change).

\begin{figure}
\includegraphics[width=15cm]{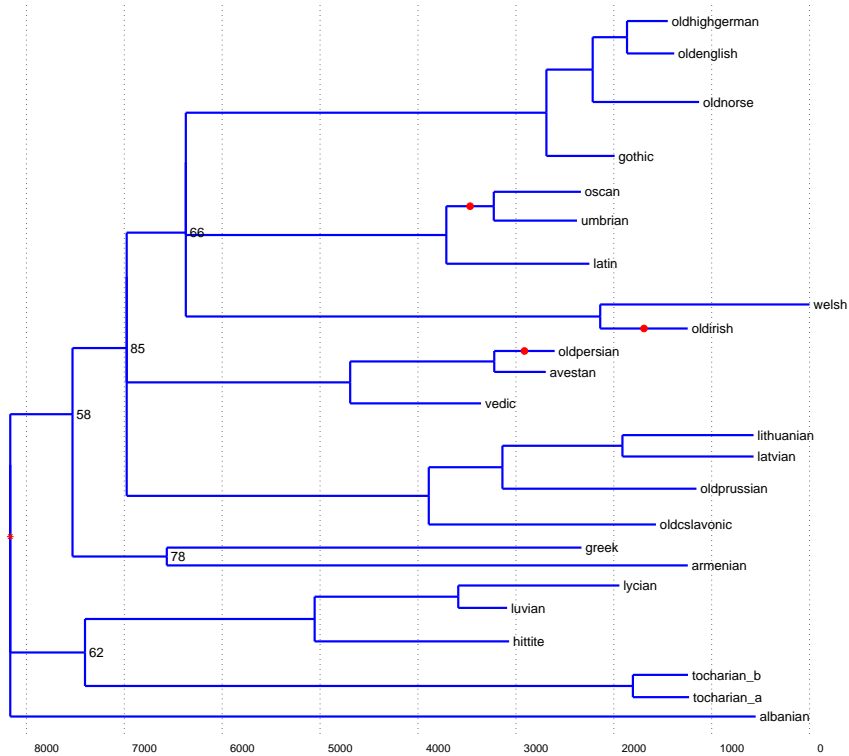}
\caption{\label{ringe} \textit{Consensus tree for the \citet{ringedata} dataset. Red dots
show catastrophes supported with probability above one half.}}
\end{figure}

We display in Figure \ref{ie8out4post} the calibration constraints on a tree sampled from the posterior. The constraints cannot be shown on a consensus tree, as slices across a consensus tree are not isochronous.

\begin{figure}
\centering
 \makebox{\includegraphics[width=16cm]{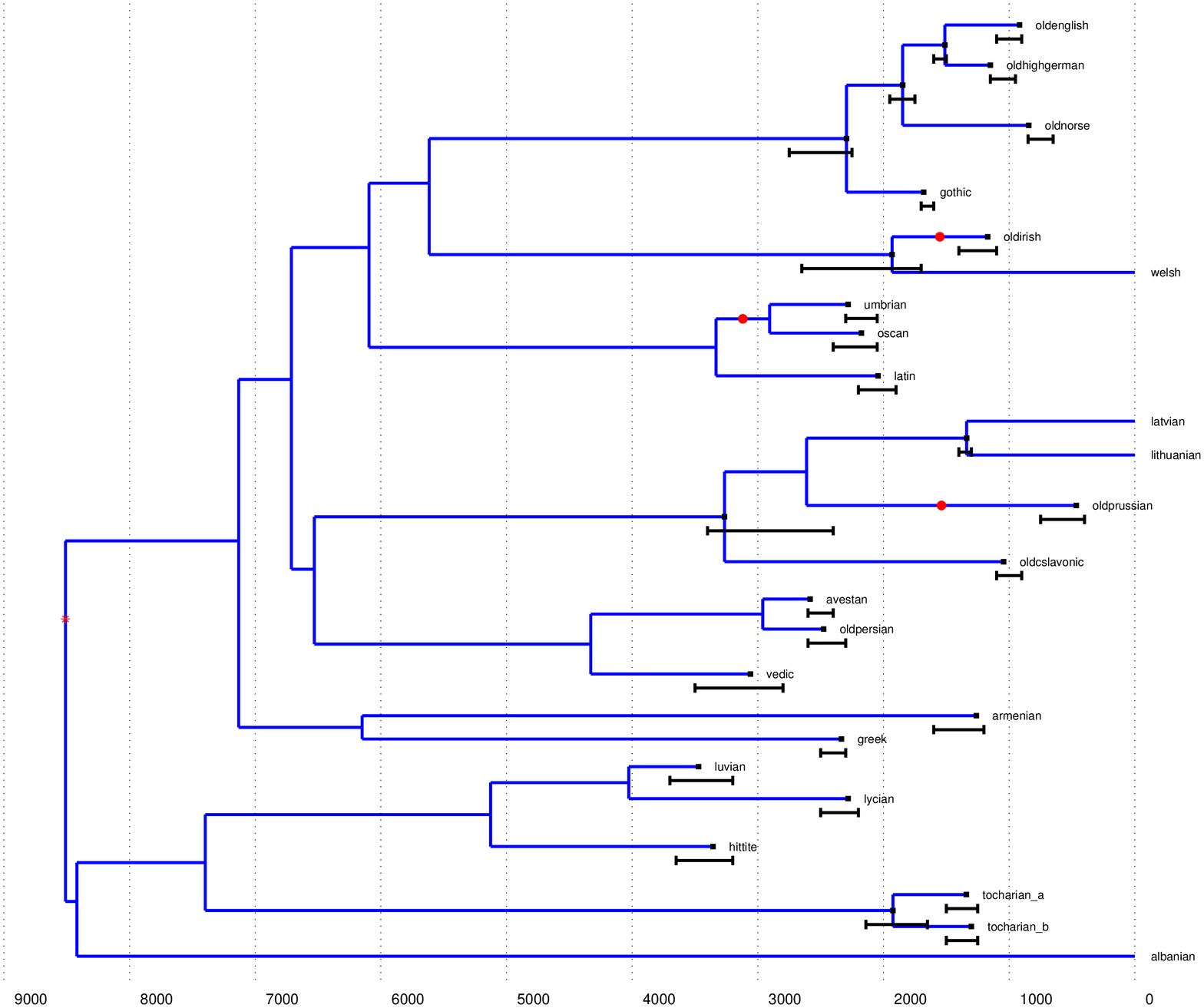}}
\caption{\label{ie8out4post}\textit{A typical sample from the posterior. All the constraints on the node ages are shown, except those for the Italic, Indo-Iranian and Iranian groups, for which we do not have an upper bound, and the root, which has an upper bound at $T=16000$ years BP.}}
\end{figure}

The analysis reconstructs some well-known features of the
Indo-European tree. The Germanic, Celtic and Italic families are
grouped together, but no particular configuration of their relative
positions is favored. The Indo-Persian group can fall outside the
Balto-Slav group but the relative position of these two is uncertain.
The deep topology of the tree is left quite uncertain by these data,
especially the position of Albanian. We find evidence for
catastrophic rate heterogeneity in three positions: on the edges
leading to Old Irish, Old Persian, and in the Umbrian-Oscan clade.


Our estimate for the root age of the Indo-European family is 8430 $\pm$ 1320 years BP. The distribution
of this key statistic is close to normal.

\end{newsec}
\begin{newsec}
\section{Conclusions}\label{discussion}

Our results give a root age for the most recent common ancestor of the 
Indo-European family of language vocabularies in agreement with earlier
phylogenetic studies. Our results are in agreement with models which 
put this date around 8500 BP, and in conflict with models which require
it to be less than 6500 years BP. Our studies of synthetic
out-of-model data, and reconstruction tests for known historical data support our view that
this main result is robust to model error. It would not be robust to a step change in the rate of
lexical diversification acting in a coordinated fashion across the
Indo-European languages extant some 3000 to 5000 years ago. 

The methods outlined here for handling missing data and rate heterogeneity in the diversification of languages, as seen through lexical data, will find applications to generic trait data.

\end{newsec}

\appendix
 
\begin{newsec}
\section{Recursions for other registration processes}\label{nounique}

This section complements Section~\ref{likelihood}: we give iterations for
$u_i^{(0)}$, $u_i^{(1)}$, $u_i^{(s_i-1)}$, $u_i^{(s_i)}$,
$v_i^{(s_i-1)}$ and $v_i^{(s_i)}$. These are the quantities
needed (as in Equation~\ref{eq:examplereg}) to
evaluate the sum in Equation~\ref{eqA}, for registration rules
which use Condition (1) in combination with other conditions
from Section~\ref{observation}.
Consider a pair of edges $\langle c_1,i\rangle$, $\langle c_2,i\rangle$ in $E$. In the notation of the text,
\begin{eqnarray*}
 u_i^{(0)} & =& \left( (1-\delta_{i,c_1})+\delta_{i,c_1}u_{c_1}^{(0)}\right) \left( (1-\delta_{i,c_2})+\delta_{i,c_2}u_{c_2}^{(0)}\right) \\
 u_i^{(1)} & = &\delta_{i,c_1}(1-\delta_{i,c_2})u_{c_1}^{(1)} + \delta_{i,c_2}(1-\delta_{i,c_1})u_{c_2}^{(1)} + \delta_{i,c_1}\delta_{i,c_2}(u_{c_1}^{(1)}u_{c_2}^{(0)} + u_{c_1}^{(0)}u_{c_2}^{(1)})\\
 u_i^{(s_i)} &=& \delta_{i,c_1}u_{c_1}^{(s_{c_1})}\delta_{i,c_2}u_{c_2}^{(s_{c_2})}\\
 u_i^{(s_i-1)}&=&\left(\delta_{i,c_1}u_{c_1}^{(s_{c_1}-1)}+\mathbb{I}_{\{s_{c_1}=1\}}(1-\delta_{i,c_1})\right)\delta_{i,c_2}u_{c_2}^{(s_{c_2})} \\
&&\quad + \delta_{i,c_1}u_{c_1}^{(s_{c_1})}\left(\delta_{i,c_2}u_{c_2}^{s_{c_2}-1}+\mathbb{I}_{\{s_{c_2}=1\}}(1-\delta_{i,c_2})\right)\qquad\\
 v_i^{(0)} &=& \left(\delta_{i,c_1}v_{c_1}^{(0)} + (1-\delta_{i,c_1})\prod_{j\in V_L^{c_1}}\xi_j\right) \left(\delta_{i,c_2}v_{c_2}^{(0)} + (1-\delta_{i,c_2})\prod_{j\in V_L^{c_2}}\xi_j\right)\\
 v_i^{(s_i)} &=& \left(\delta_{i,c_1}v_{c_1}^{(s_{c_1})}+(1-\delta_{i,c_1})\prod_{j\in V_L^{c_1}}(1-\xi_j)\right) \left(\delta_{i,c_2}v_{c_2}^{(s_{c_2})}+(1-\delta_{i,c_2})\prod_{j\in V_L^{c_2}}(1-\xi_j)\right)\\
 v_i^{(s_i-1)} &=& \left(\delta_{i,c_1}v_{c_1}^{(s_{c_1}-1)}+(1-\delta_{i,c_1})s_{c_1}(1-\xi)^{s_{c_1}-1}\xi\right) \left(\delta_{i,c_2}v_{c_2}^{(s_{c_2})}+(1-\delta_{i,c_2})(1-\xi)^{s_{c_2}}\right) \nonumber \\
&& +\ \left(\delta_{i,c_1}v_{c_1}^{(s_{c_1})}+(1-\delta_{i,c_1})(1-\xi)^{s_{c_1}}\right) \left(\delta_{i,c_2}v_{c_2}^{(s_{c_2}-1)}+(1-\delta_{i,c_2})s_{c_2}(1-\xi)^{s_{c_2}-1}\xi\right)
\end{eqnarray*}
The recursion is evaluated from the leaves $i\in V_L$, at which
\begin{eqnarray*}
u_{i}^{(0)}&=&u_i^{(s_i-1)}=1-\xi_i\\
u_{i}^{(1)}&=&u_i^{(s_i)}=\xi_i\\
v_i^{(0)}&=&v_i^{(s_i-1)}=0\\
v_i^{(s_i)}&=&1 
\end{eqnarray*}
\end{newsec}

\bibliography{biblio}

\end{document}